\documentclass[tradiabstract]{aa}  

\usepackage{graphicx}
\usepackage{txfonts}
\usepackage[dvipsnames]{xcolor}
\usepackage{natbib,twoopt}
\usepackage[breaklinks=true]{hyperref} 
\bibpunct{(}{)}{;}{a}{}{,} 
\makeatletter
\newcommandtwoopt{\citeads}[3][][]{\href{http://adsabs.harvard.edu/abs/#3}%
{\def\hyper@linkstart##1##2{}%
\let\hyper@linkend\@empty\citealp[#1][#2]{#3}}}
\newcommandtwoopt{\citepads}[3][][]{\href{http://adsabs.harvard.edu/abs/#3}%
{\def\hyper@linkstart##1##2{}%
\let\hyper@linkend\@empty\citep[#1][#2]{#3}}}
\newcommandtwoopt{\citetads}[3][][]{\href{http://adsabs.harvard.edu/abs/#3}%
{\def\hyper@linkstart##1##2{}%
\let\hyper@linkend\@empty\citet[#1][#2]{#3}}}
\newcommandtwoopt{\citeyearads}[3][][]%
{\href{http://adsabs.harvard.edu/abs/#3}
{\def\hyper@linkstart##1##2{}%
\let\hyper@linkend\@empty\citeyear[#1][#2]{#3}}}
\makeatother

\begin{document} 

   \title{The IBISCO survey}

   \subtitle{I. Multiphase discs and winds in the Seyfert galaxy Markarian 509}

   \author{M.V. Zanchettin\inst{\ref{inst7}, \ref{inst1},\ref{inst2}} \and C. Feruglio\inst{\ref{inst2}, \ref{inst8}} \and M. Bischetti\inst{\ref{inst2}}  \and A. Malizia\inst{\ref{inst3}} \and M. Molina\inst{\ref{inst3}} \and A. Bongiorno\inst{\ref{inst4}} \and M. Dadina\inst{\ref{inst3}} \and C. Gruppioni\inst{\ref{inst3}} \and E. Piconcelli\inst{\ref{inst4}} \and F. Tombesi\inst{\ref{inst5},\ref{inst4},\ref{inst6},\ref{inst9}} \and A. Travascio\inst{\ref{inst2}} \and F. Fiore\inst{\ref{inst2}, \ref{inst8}}}

   \institute{SISSA, Via Bonomea 265, I-34136 Trieste, Italy \email{mazanch@sissa.it}\label{inst7} \and Dipartimento di Fisica, Sezione di Astronomia, Universita degli Studi di Trieste, Via Tiepolo 11, I-34143 Trieste, Italy\label{inst1} \and INAF Osservatorio Astronomico di Trieste, via G.B. Tiepolo 11, 34143 Trieste, Italy\label{inst2} \and IFPU - Institute for fundamental physics of the Universe, Via Beirut 2, 34014 Trieste, Italy \label{inst8} \and INAF-OAS Bologna, via Gobetti 101, I-40129 Bologna, Italy\label{inst3} \and INAF - Osservatorio Astronomico di Roma, Via di Frascati 33, 00040, Monteporzio Catone, Rome, Italy\label{inst4}    \and Dept. of Physics, University of Rome ‘Tor Vergata’, via della Ricerca Scientifica 1, 00133, Rome, Italy\label{inst5} \and Dept. of Astronomy, University of Maryland, College Park, MD, 20742, USA\label{inst6} \and NASA - Goddard Space Flight Center, Code 662, Greenbelt, MD 20771, USA\label{inst9} }

   \date{Received 2020 October 27 ; accepted 2021 July 13 }

 
    \abstract
   {We present the analysis of the ALMA CO(2-1) emission line and the underlying 1.2 mm continuum of Mrk509 with spatial resolution of $\sim$270 pc. This local Seyfert 1.5 galaxy, optically classified as a spheroid, is known to host a ionised disc, a starburst ring, and ionised gas winds on both nuclear (Ultra Fast Outflows) and galactic scales. 
   From CO(2-1) we estimate a molecular gas reservoir of $M_{H_2} = 1.7 \times 10^9\, \mathrm{M_{\odot}}$, located within a disc of size $\sim$5.2 kpc, with $M_{dyn}$ = (2.0$\pm$1.1) $\times$ $10^{10}$ $\mathrm{M_{\odot}}$ inclined at $44\pm10$ deg. The molecular gas fraction within the disc is $\mu_{gas} = 5 \%$, consistent with that of local star-forming galaxies with similar stellar mass. The gas kinematics in the nuclear region within r$\sim$700 pc, that is only marginally resolved at the current angular resolution, suggests the presence of a warped nuclear disc. 
   Both the presence of a molecular disc with ongoing star-formation in a starburst ring, and the signatures of a minor merger, are in agreement with the scenario where galaxy mergers produce gas destabilization, feeding both star-formation and AGN activity.
  The spatially-resolved Toomre Q-parameter across the molecular disc is in the range $Q_{gas}=0.5-10$, and shows that the disc is marginally unstable across the starburst ring, and stable against fragmentation at nucleus and in a lopsided ring-like structure located inside of the starburst ring. 
   We find complex molecular gas kinematics and significant kinematics perturbations at two locations, one within 300 pc from the nucleus, and one 1.4 kpc away close to the region with high $Q_{gas}$, that we interpret as molecular winds with velocity $v_{98} = 200-250 \, \rm km/s$. The total molecular outflow rate is in the range 6.4 - 17.0 $\mathrm{M_\odot/yr}$, for the optically thin and thick cases, respectively. 
   The molecular wind total kinetic energy is consistent with a multiphase momentum-conserving wind driven by the AGN with $\dot{P}_{of}/\dot{P}_{rad}$ in the range 0.06-0.5 .
   The spatial overlap of the inner molecular wind with the ionised wind, and their similar velocity suggest a cooling sequence within a multiphase wind driven by the AGN. The second outer molecular wind component overlaps with the starburst ring, and its energy is consistent with a SN-driven wind arising from the starburst ring.}  


   \maketitle
%

\section{Introduction}

Understanding the relation between the Feedback from accreting super-massive black holes (SMBHs) and the interstellar medium (ISM) of their host galaxies is still an open issue. Host bulge properties such as velocity dispersion, luminosity and mass are tightly correlated with the mass of the SMBHs in the galactic center as shown by past seminal works \citep{gebhardt,ferrareseford,kormendyho,shankar2016,shankar2017}. The gas in the galactic bulge inflows toward the nucleus and gives rise to the growth of central SMBH through the active luminous phase of the Active Galactic Nuclei (AGN) \citep{fabian2012,kingpounds}. During their active phases, AGN can generate winds that interact with galaxy ISM, potentially altering both the star formation process and the nuclear gas accretion. When the black hole reaches a critical mass the AGN driven winds, the nuclear activity and the SMBH growth are stopped giving rise to the SMBH - host bulge properties relations \citep{silkrees,fabian1999, king2003}. \\ Outflows are ubiquitous in both luminous AGN and in local Seyfert galaxies, and occur on a wide range of physical scales, from highly-ionised semi-relativistic winds and jets in the nuclear region at sub-parsecs scales, out to galactic scale outflows seen in mildly ionised,  molecular and neutral gas  \citep[][and references therein]{morganti2016, fiore2017, fluetsch2019, lutz2020, veilleux2020}. In some cases molecular and ionised winds have similar velocities and are nearly co-spatial, suggesting a cooling sequence scenario where molecular gas forms from the cooling of the gas in the ionised wind \citep{richings2017, menci2019}. Other AGNs show ionised winds that are faster than the molecular ones, suggesting a different origin of the two phases \citep[][and references therein]{veilleux2020}. 
The molecular phase is a crucial element of the feeding and feedback cycle of AGN because it constitutes the bulk of the total gas mass and it is the site of star formation activity. On galactic scales, massive molecular winds are common in local Seyfert galaxies \citep[e.g.][]{feruglio2010,cicone2014,dasyra2014,morganti2015,garcia-burillo2014,garcia-burillo2017,garcia-burillo2019}, these winds likely suppress star formation (i.e. negative feedback) as they reduce the molecular gas reservoir by heating or expelling gas from the host-galaxy ISM. In late type AGN-host galaxies, the gas kinematics appears complex at all scales, showing several components such as bars, rings, and (warped) discs, with high velocity dispersion regions  \citep[e.g.][]{shimizu,feruglio2020,fernandez2020,alonso-herrero2020,aalto2020,audibert2020}. 
Accurate dynamical modeling of the molecular gas kinematics reveals kinematically-decoupled nuclear structures, high velocity dispersion at nuclei, trailing spirals, evidences of inflows and AGN-driven outflows.  \citep[e.g.][]{combes2019a,combes2019b,combes2020}.
The outflows driving mechanism (wind shock, radiation pressure or jet), their multiphase nature and their relative weights and impact on the galaxy ISM are still open problems \citep{fauchergigure2012,zubovas2012,richings2017,menci2019,ishibashi2021}.
So far different outflow phases have been observed only for a handful of sources. Atomic, cold and warm molecular outflows have been observed in radio galaxies \citep[e.g.][]{morganti2007,dasyracombes2012, dasyra2014, tadhunter2014, osterloo2017}.
The nuclear semi-relativistic phase and the galaxy scale molecular phase have been observed simultaneously in less than a dozen sources, with varied results: in some cases data suggest energy driven flows \citep{feruglio2015,tombesi2015, longinotti2018, smith2019}, in other cases data suggest momentum driven flows \citep[e.g.][]{garcia-burillo2014,feruglio2017,fluetsch2019,bischetti2019, marasco2020}.  
\citet{fiore2017}, using a compilation of local and high redshift winds, showed that there is a broad distribution of the momentum boost, suggesting that both energy and momentum-conserving expansion may occur. Enlarging the sample of local AGN-host galaxies with outflows detected in different gas phases is important to understand the nature and driving mechanisms of galaxy-scale outflows.\\ 
In this paper we present a study of Markarian 509 (Mrk 509) based on ALMA observations. 
Mrk 509 is a Seyfert 1.5 galaxy located at D $\sim$ 142.9 Mpc \citep[z = 0.034397,][]{hucra} implying a physical scale of 0.689 kpc/arcsec, and hosts an AGN with $L_{2-10 keV} = 10^{44.16}\mathrm{\,erg\,s^{-1}}$ \citep{shinozaki2006}, a bolometric luminosity of $L_{Bol}=  10^{44.99}\mathrm{\,erg\,s^{-1}}$ \citep{duras2020}, and a black hole with mass $M_{bh} = 10^{8.04}~\mathrm{M_{\odot}} $ \citep{bentz}.  
The AGN is hosted by a medium size galaxy resembling a bulge in optical imaging \citep{Ho2014}, but hosting both an ionized gas disc and a starburst ring \citep{phillips1983, kriss2011, fisher}, where star formation is currently ongoing \citep[$SFR\sim 5 \rm M_{\odot}/yr$,][]{gruppioni2016}.
 
 \noindent Mrk 509 is a complex system showing evidence of both an ongoing minor merger and multiphase gas winds. The minor merger with a gas-rich dwarf is suggested by an ionised gas linear tail \citep{fischer2013, fisher}, whereas \citet{liu} found evidence of a [OIII] quasi-spherical wind powered by the AGN in the inner $\sim 2$ kpc region, with velocity 290 km/s and mass outflow rate $\dot{M}_{of}$ = 5 $\mathrm{M_{\odot}/yr}$. The ionised wind appears spatially anti-correlated with the $H\beta$ emission, suggesting suppression of star formation in this region.
Gas kinematics suggests that the ionised wind is physically unrelated to the linear tail, and their apparent overlap is due to projection effects only \citep{liu}.
On nuclear scales, highly ionised semi-relativistic gas winds are detected with velocity v$\sim$ 0.15-0.2c and outflow rates in the range 0.005-0.05 $\mathrm{M_{\odot}/yr}$ \citep{dadina2005,cappi2009,tombesi2011, tombesi2012,detmers}. 
Simultaneous Chandra and Hubble Space Telescope with the Cosmic Origins Spectrograph (HST-COS) observations showed that the wind detected through UV absorption lines is located within $\sim200$ pc from the active nucleus, and that X-ray warm absorbers and UV absorbers are related \citep{ebrero2011,kriss2019}. 
Given the widespread evidence of winds from nuclear to kpc scales and in different gas phases \citep{phillips1983,kriss2011,tombesi2011, tombesi2012,liu}, Mrk 509 is an optimal candidate to investigate the wind driving mechanisms, stellar feedback versus AGN feedback, and their impact on the host-galaxy ISM.
In this paper, we exploit 
ALMA data of Mrk509 to trace the distribution and kinematics of cold molecular gas and to look for evidences of molecular winds. 
The CO(2-1) and underlying continuum observations examined here are part of the ALMA survey of the IBISCO sample.
 The IBISCO sample is drawn from the  INTEGRAL (International Gamma-Ray Astrophysics Laboratory) IBIS (Imager on Board INTEGRAL Spacecraft) catalog of hard X-ray (20-100 keV) selected AGN \citep{malizia}. The IBISCO sample is selected to have z$<$0.05, AGN bolometric luminosity $L_{Bol}>10^{43}\mathrm{\,erg\,s^{-1}}$, and accurate black hole mass measurements. Being hard X-ray selected, the IBISCO sample is unbiased against nuclear obscuration, similarly to BASS, the BAT AGN Spectroscopic Survey \citep{koss2017, koss2020}. 
The ALMA survey of IBISCO is aimed at obtaining a reliable overview of the cold gas kinematics in the central $\sim $ kpc of the AGN host galaxies with angular resolutions in the range 100-200 pc. 
\newline \noindent The paper is organized as follows. Section 2 presents the ALMA observational setup and data reduction. Section 3 presents the observational results, in particular the continuum emission, the gas properties and the CO kinematics. Section 4 describes the dynamical modeling, section 5 the properties of the molecular wind. In Section 6 we discuss our results and in section 7 we show our conclusions and future perspectives.

\begin{figure*}[h!]
  \resizebox{\hsize}{!}{\includegraphics{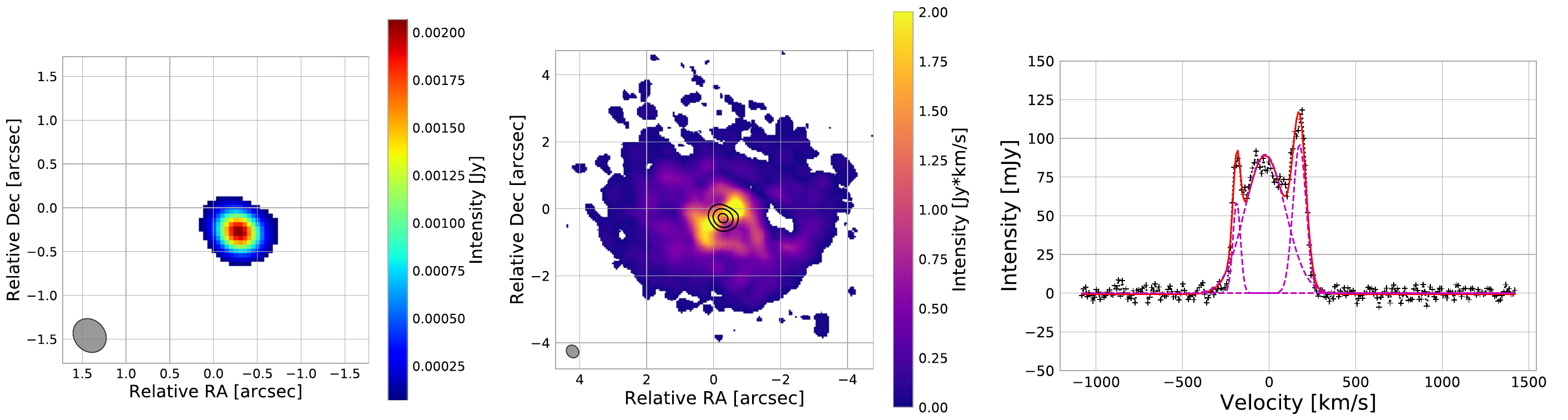}}
  \caption{Left panel: 1.2 mm continuum map of Mrk509. Regions with emission below 5$\sigma$ have been blanked. The synthesized beam (0.396$ \times$ 0.343 $\mathrm{arcsec^2}$) is shown by the grey filled ellipse. Central panel: CO(2-1) integrated flux map,  where a mask with threshold of $3\sigma $} has been applied. Black contours show the 1.2 mm continuum emission at (5,30,100)$\times \sigma$. The synthesized beam is reported bottom left corner. Right panel: CO(2-1) spectrum (black crosses) extracted from the continuum-subtracted clean data-cube from the regions where emission is above 3$\sigma$. The solid red shows the multi-gaussian fit, the three gaussian components are shown using magenta lines.
  \label{spettro}
\end{figure*}

\begin{figure*}
   \centering
   \resizebox{\hsize}{!}{\includegraphics{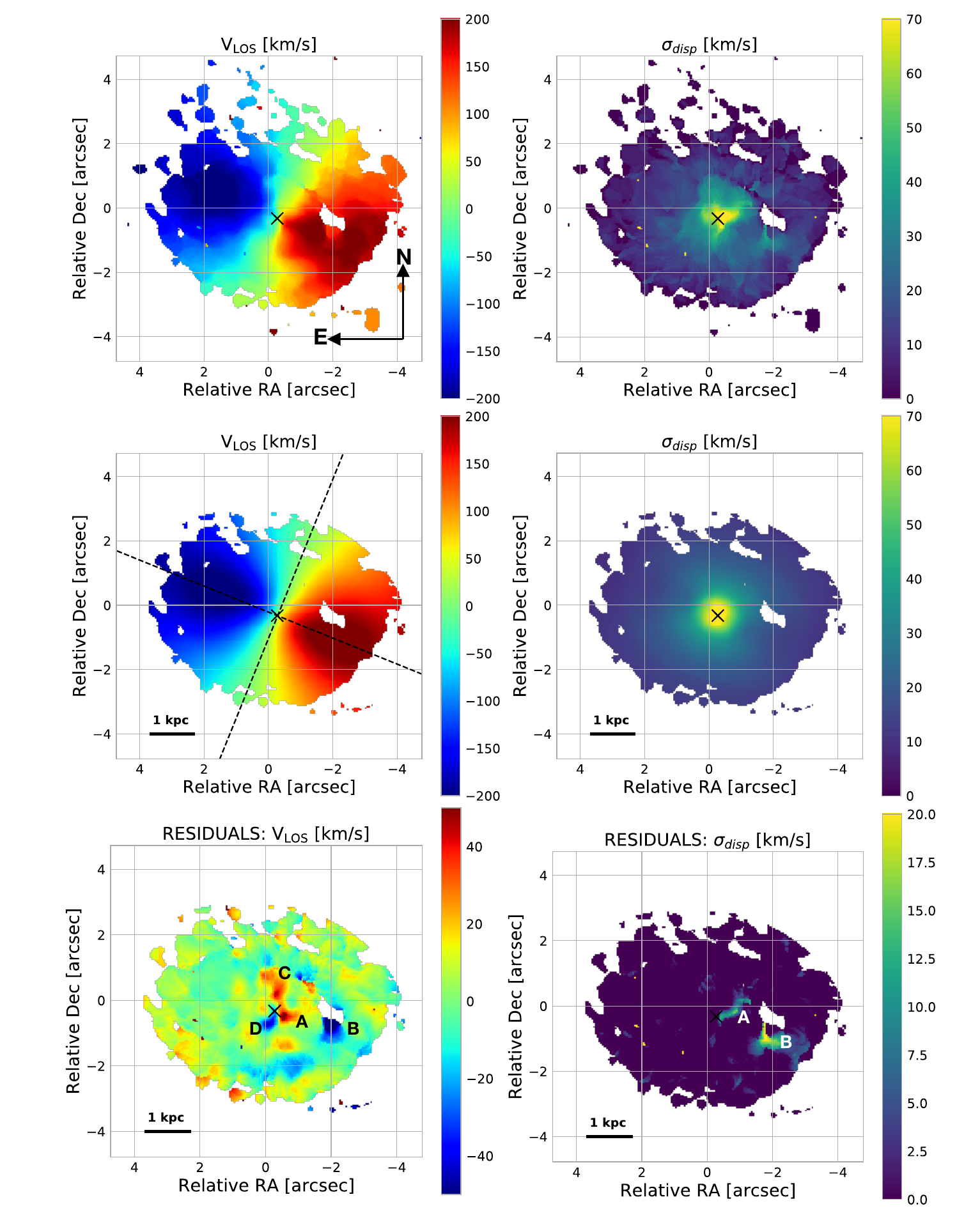}}
   \caption{Top panels: the CO(2-1) moment 1 (top-left) and moment 2 maps (top-right). A threshold of $3\sigma $ has been applied. The black cross marks the AGN position. Central panels: The moment 1 (central-left) and moment 2 (central-right) maps of the  $^{3D}BAROLO$ disc model. The black dashed lines mark the orientation of the major and minor kinematic axis (PA=248 and 338 deg, respectively). Bottom panels: the moment 1 (bottom-left) and moment 2 (bottom-right) maps of the residuals obtained by subtracting the model velocity and velocity dispersion maps from the observed ones.  A, B, C and D mark the main dynamical perturbations discussed in the text.}
              \label{data-model-res}
\end{figure*}

\section{Observations and data reduction}

Mrk 509 was observed with ALMA (program ID 2017.1.01439.S) during cycle 5 on September 2018 for a total integration time of 75 minutes in the frequency range 221.8-223.7 GHz (band 6), which covers the CO(2-1) line (observed redshifted frequency of 222.9 GHz) and the underlying 1.2 mm continuum. The obervations were conducted with the 12 m array in the
C43-5 configuration with 44 antennas, minimum baseline of 14.6 m and maximum baseline of 1397.9 m. This provided us an angular resolution of about 0.4 arcsec and a largest angular scale of $\sim$ 4.6 arcsec.\\ We used CASA 5.4.1 software \citep{mcmullin} to produce a set of calibrated visibilities and to generate clean data-cubes. We calibrated the visibilities in the pipeline mode and using the default calibrators provided by the observatory: bandpass and flux calibrators J2148+0657 and J2000-1748, J2025-0735 phase/amplitude calibrator. The absolute flux accuracy is better than 10$\%$. To estimate the rest-frame 1.2 mm continuum emission we averaged the visibilities in the four spectral windows, excluding the spectral range covered by the CO(2-1) emission line. To estimate the continuum emission underlying the CO(2-1) line, we modeled the visibilities in the spectral window (spw) containing the emission line with the \texttt{uvcontsub} task, adopting a first order polynomial. We verified that using a zero order polynomial did not affect the results. We subtracted the continuum from the global visibilities and created the continuum-subtracted CO(2-1) visibility table. We imaged CO(2-1) and continuum using the \texttt{tclean} task by adopting a natural weighting scheme, and produced a clean cube by using the \texttt{hogbom} cleaning algorithm with a detection threshold of 3 times the r.m.s. noise, a pixel size of 0.05 arcsec and a spectral resolution of 10.6 km/s. With this procedure we obtained a clean map of the continuum with a synthesized beam of 0.396 $\times$ 0.343 $\mathrm{arcsec^2}$ at PA = 51 deg, which corresponds to a spatial resolution of $\sim$ 285 $\times$ 244 $\mathrm{pc^2}$. The resulting r.m.s. noise in the clean continuum map is 0.014 mJy/beam in the aggregated bandwidth. From the continuum-subtracted CO(2-1) visibilities  we produced a clean data-cube with a beam size of 0.412 $\times$ 0.353 $\mathrm{arcsec^2}$ at a PA = 48 deg and a r.m.s. noise of 0.38 mJy/beam for a channel width of 10.6 km/s.

\section{Results}\label{sec:results} 

Figure \ref{spettro}, left panel, shows the 1.2 mm continuum map. The peak position of the continuum flux density map, 
obtained through 2D fitting in the image plane with CASA, is consistent with the AGN position reported by NED (NASA/IPAC Extragalactic Database, https://ned.ipac.caltech.edu). We measure a continuum flux density $S_{1.2 mm} = 2.17 \pm 0.04$ mJy, and a flux at the peak position $S_{peak} = 2.09 \pm 0.02$ mJy/beam. According to our 2D fit the continuum emission is consistent with an unresolved source.

\begin{table}
     \caption[]{Parameters of the Gaussian fit of the CO(2-1) line.}
         \label{tab:fit-spettro}
\centering                          
\begin{tabular}{c c c }        
\hline\hline                 
FWHM [km/s] & $I_{Peak}$ [mJy] & Velocity [km/s]  \\
(1) & (2) & (3)  \\
\hline                        
            21.60 $\pm$ 0.03   &  59.4 $\pm$0.1  & -187.71 $\pm$ 0.03    \\
            116.82 $\pm$ 0.08   &  89.5 $\pm$ 0.3  & -23.56 $\pm$ 0.05    \\
            37.63 $\pm$ 0.02   &  96.2 $\pm$ 0.1  & 175.39 $\pm$0.02    \\
\hline  
\end{tabular}\\
  \flushleft 
 \footnotesize{ {\bf Notes.} Parameters of the three Gaussian components plotted in Fig. 1 (1) the FWHM, (2) the intensity at the peak in mJy, (3) the velocity at the peak w.r.t the rest frame. Errors are given at 1$\sigma$.}
\end{table}

\noindent Figure \ref{spettro}, central panel, shows that the CO(2-1) surface brightness (moment 0 map) is distributed over a region of approximately 8 arcsec, that is 5.5 kpc, and, in the inner $\sim$1.4 kpc region, shows an elongated shape along the north-west to south-east direction, crossing the AGN position. A CO-depleted region, that is a region  with a CO(2-1) surface brightness lower than neighbouring regions, is detected located at 1.4 kpc west-ward to the AGN. 
The CO(2-1) spectrum (Fig. \ref{spettro}, right panel), extracted from the clean data-cube in the region defined by a mask with 3$\sigma$ threshold, shows a multiple peak profile reminiscent of disc-like kinematics. We fitted the emission line with a combination of three Gaussian profiles whose parameters are reported in the Table \ref{tab:fit-spettro}. From the best-fit we measured an integrated flux density of $S_{CO} = 36.7 \pm \mathrm{0.8 \, Jy\, km/s}$ over a line width of 180 km/s. We then derived the CO(2-1) luminosity $L'_{CO(2-1)}= (5.0 \pm 0.1)\times 10^8$ $\rm \mathrm{K\,km/s\,pc^2}$ using the relation of \citet{solomon}.  
 To estimate the proper conversion factor $\alpha_{CO}$, we use the stellar mass $M_*$ and the conversion function by \citet{accurso}. 
The $M_*$ is derived from  broad-band spectral energy distribution (SED) decomposition including the emission of stars, dust heated by SF and AGN dusty torus \citep{gruppioni2016}. We report in Appendix \ref{appendixa} details on the SED fitting. We derive a total stellar mass of $M_* = 1.2 \times 10^{11} \, \mathrm{M_{\odot}}$, and a $\alpha_{CO}\,= 3.3\,\mathrm{M_{\odot} (K\,km/s\,pc^{2})^{-1}}$.
Using this, we derive a molecular gas mass $M_{H_2} = (1.65\, \pm \, 0.04)\times 10^9\, \mathrm{M_{\odot}}$ within a region of $\sim 5.2$ kpc diameter.
For the CO-depleted region we derived a $3\sigma$ upper limit molecular gas mass $M_{H_2} < 0.35 \times 10^7\mathrm{M_{\odot}}$ assuming the same conversion factor.
\\ The mean-velocity map (moment 1 map, Fig. \ref{data-model-res} top-left panel) shows a gradient oriented north-east to south-west along a position angle (PA) of $\sim$ 250 deg, with range from -200 to 200 km/s, which likely indicates an inclined rotating disc (PA in degrees is measured anti-clockwise from North).
 Figure \ref{data-model-res} (top-right panel) shows the CO(2-1) velocity dispersion (moment 2 map), $\sigma_{disp}$, with values of 10-20 km/s in the outer regions. In the inner 1.4 $\times$ 1.4 ${\rm kpc^2}$ (corresponding to 2$\times$2 $\mathrm{arcsec^2}$), $\sigma_{disp}$ has a butterfly-shaped morphology with a $\sigma_{disp}$ = 80 km/s towards the central beam, where beam-smearing effects can boost the $\sigma_{disp}$ \citep[e.g][]{davies2011}. A more reliable estimate of the real $\sigma_{disp}$ in the nuclear region is 50-60 km/s, as measured in an annulus with 0.3 < R < 0.6 arcsec. We detect a region with enhanced $\sigma_{disp}$ = 30-40 km/s located south of the CO-depleted region, suggesting that also at this location dynamic perturbations are likely present. 
\begin{figure}
  \resizebox{\hsize}{!}{\includegraphics{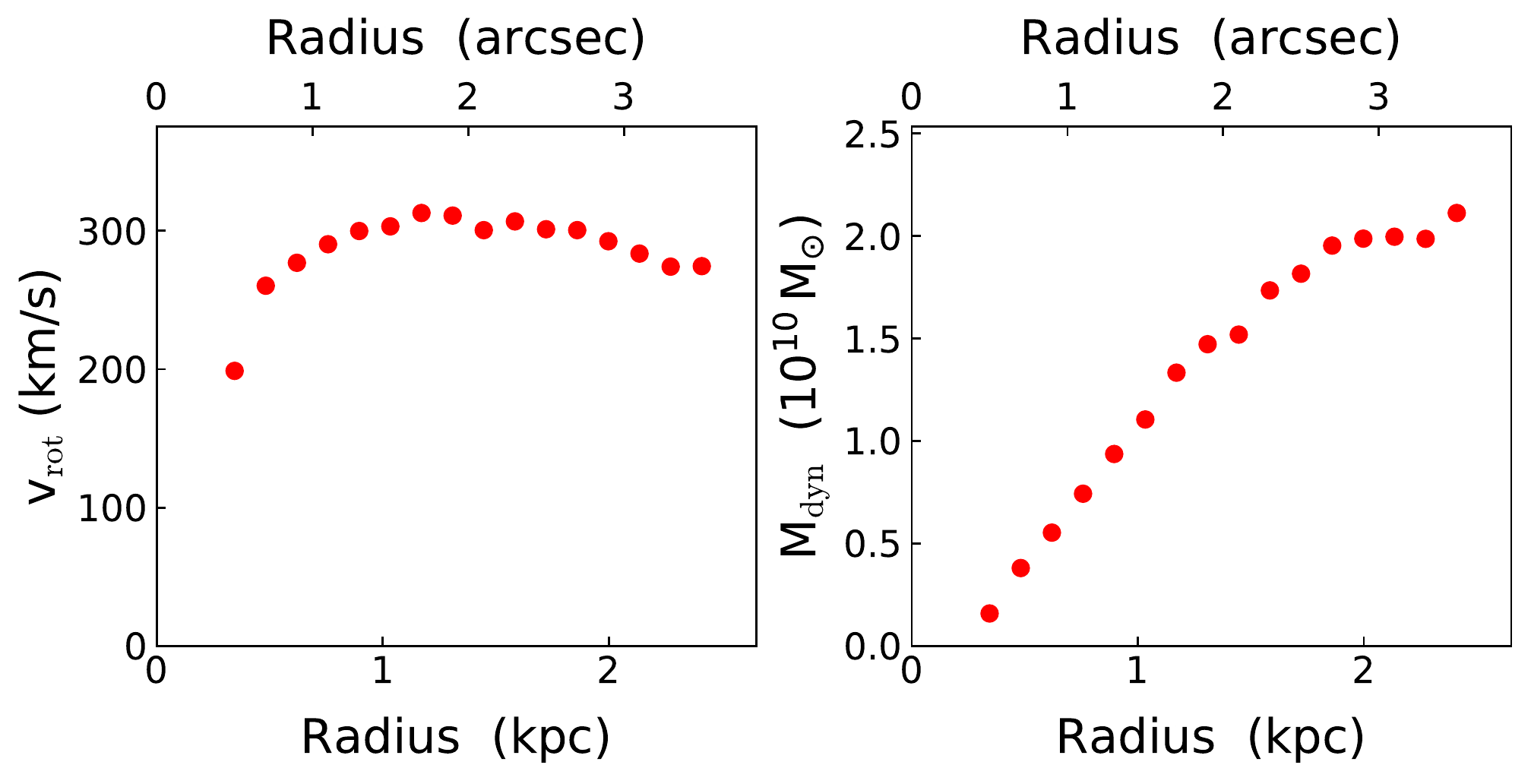}}
  \caption{Left panel: rotational velocity $v_{rot}$ versus radius of the best fit disc model of the main disc. Right panel: dynamical mass versus radius within the inner $\sim$2.5 kpc radius, derived from the relation $M_{dyn} = rv^2_{rot}/2G$.}
  \label{parameters}
\end{figure}
\begin{figure*}
   \centering
   \resizebox{\hsize}{!}{\includegraphics{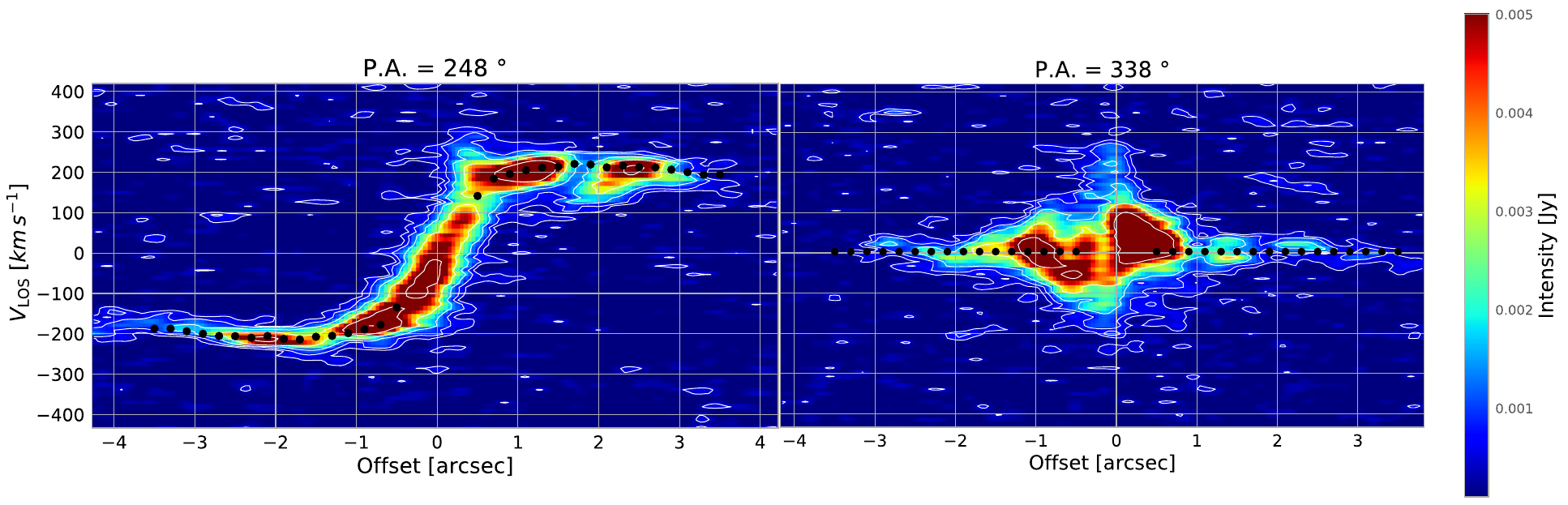}}
   \caption{Left panel: PV plot along PA = 248 deg ( i.e. the kinematic major axis). Right panel: PV plot along PA = 338 deg (i.e. the kinematic minor axis). The slit width is set to 0.45 arcsec (i.e. approximately the FWHM size of the synthetic beam major axis). The white contours refer to data, contours are drawn at (1,2,4,8,16,32)$\times \sigma$, black dots mark the $V_\mathrm{LOS}$ value from $^{3D}BAROLO$ model.}
              \label{pv}
    \end{figure*}

\section{Molecular Discs}

Based on the moment maps we build a dynamical model of the system using $^{3D}BAROLO$ \citep["3D-Based Analysis of Rotating Objects from Line Observations”,][]{diteodoro}. 
Any deviation from a rotating disc kinematics can then be identified by comparing the disc dynamical model with the observed data, 
through the residual data-cube, that is obtained by subtracting the model from the observed data-cube. 
We fitted a 3D tilted-ring model to the emission line data-cube to provide a first order dynamical model of the CO-emitting gaseous disc. In the first run, we allowed four parameters to vary: rotation velocity, velocity dispersion, the disc inclination and position angle. We fixed the kinematic center to the 1.2 mm continuum position (Sec. \ref{sec:results}). 
We then run the model again fixing the inclination, so only PA, rotation velocity and velocity dispersion are allowed to vary. The inclination from the minor/major axis ratio from UV imaging (GALEX archive) is $44\pm10$ deg, and we fix it to 44 deg in the model.
The first run produces residuals of amplitude about 5\% in the moment 0 map, that further increase by adding a radial velocity component. The second run, with fixed inclination, produces residuals of amplitude about 2\%, therefore we adopt the latter as best-fit disc model. 
The disc model has a position angle PA = $248 \pm 5$ deg, a $v_{LOS}$ that ranges from -200 to 200 km/s and a velocity dispersion that decreases from 70 km/s in the inner region to $\sim$10 km/s in the outer part of the disc (Fig. \ref{data-model-res}, central panels). 
Fig.\ref{parameters} shows the rotation velocity (left panel), of the best fit disc model, and the virial dynamical mass (right panel) $M_{dyn} = rv^2_{rot}/2G$ as a function of the radius out to $\sim$ 2.5 kpc. 
The rotation velocity shows a maximum value of $\sim$ 300 km/s and the dynamical mass enclosed within the maximum radius is $\sim 2.1 \times 10^{10}\rm M_\odot$.
 \newline \noindent We computed the residual velocity and velocity dispersion maps by subtracting the disc model from the data (Fig. \ref{data-model-res}, bottom panels). Residual maps indicate that the model nicely describes the observed large-scale velocity gradient, from r$\sim$1 arcsec to the outer boundary. Dynamical perturbations are detected in the residual velocity map (Fig. \ref{data-model-res}, bottom left panel) at  positions A, C (redshifted), and  B, D  (blueshifted). In the velocity dispersion residual map (Fig. \ref{data-model-res}, bottom right panel), residuals are present only at positions A and B.
The position-velocity (PV) plots shown in Fig. \ref{pv} capture the main kinematics components seen in the moment maps and in the residual maps (Fig.\ref{data-model-res}). The slices are taken through the continuum peak position, using a slit 0.45 arcsec wide (i.e. approximately the FWHM size of the synthetic beam major axis). 
The PV diagram along the kinematic major axis (PA=248 deg) shows  kinematic perturbations at offset 0.5 arcsec with $V_{LOS}$ $\sim$ 250 km/s (A), and at offset 2 arcsec with $V_{LOS}$ $\sim$ 150 km/s (B), on top of the typical disc rotation pattern. 
The PV diagram along the kinematic mainor axis (PA = 338 deg) shows both blue-shifted and red-shifted non-rotational motions with velocities that reach 300 km/s and  $-200$ km/s in the inner $\pm$ 0.5 arcsec region (C,D). 
\begin{figure*}
   \centering
   \resizebox{\hsize}{!}{\includegraphics{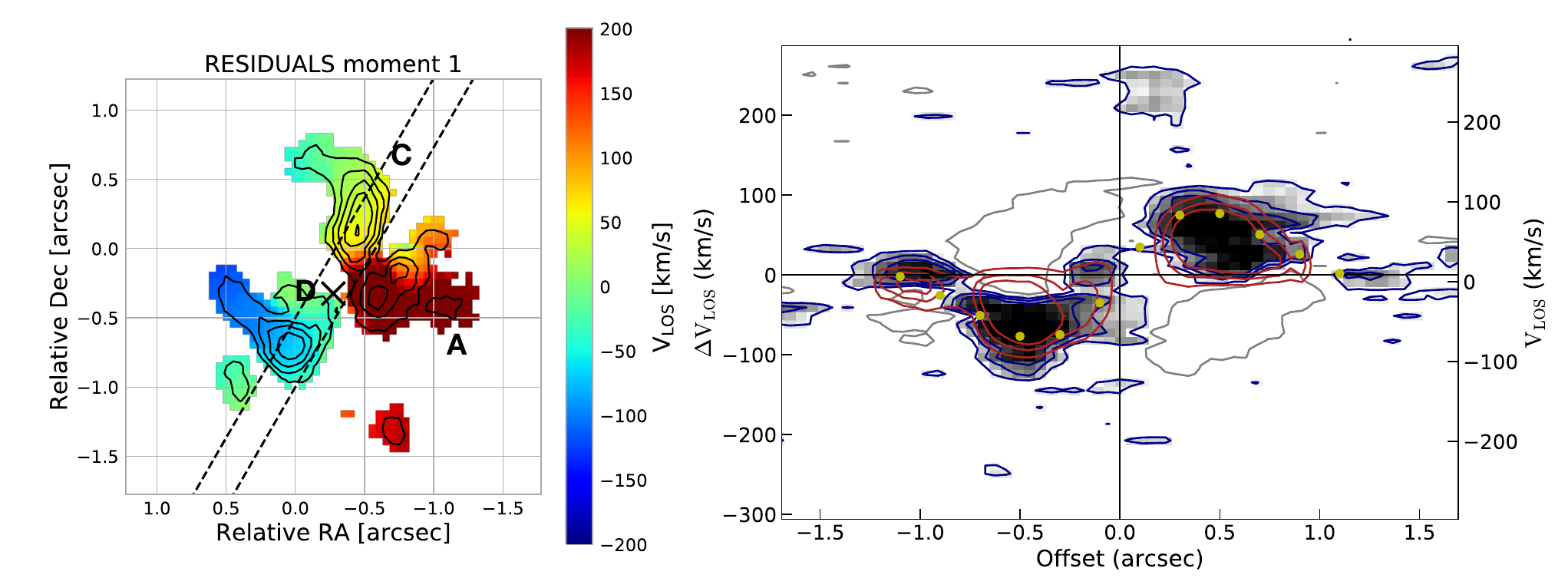}}
   \caption{Left panel: The velocity map generated from the residual cube obtained subtracting the disc model from the data-cube (a mask with a 2$\sigma$ threshold on the residual intensity map has been applied). The black contours show the CO(2-1) emission in residual intensity map at (3, 4, 5, 6, 7)$\times \sigma$. The black dashed lines represent the slice adopted for the PV diagram (right panel) and comprise the two main kinematic perturbations in the residual intensity map, along PA = 330 deg. The black cross marks the AGN position. Right panel: PV diagram along PA = 330 deg (kinematic major axis). Red contours and yellow points represent the disc model of this inner region, blue contours represent the data (contours are drawn at (2,4,6)$\times \sigma$). A, C and D mark the main dynamical perturbations.} 
              \label{residuo0}
\end{figure*}
\noindent  The perturbations C and D suggest the presence of a warped disc or bar. 
The nuclear region is barely resolved at the current angular resolution. Here we attempt to model it with a nuclear disc, as detailed in Appendix \ref{appendixb}. The modeling with a disc of inclination fixed to 66 deg and PA fixed to 330 deg, i.e. along the two brightest emission regions in the residual intensity map (C and D, black contours in Fig. \ref{residuo0} left panel), is able to account for and nicely model the residual emission (Fig. \ref{residuo0}, right panel, and Fig. \ref{outflow}, top panel), suggesting the presence of a nuclear warped disc.
We note that observations with higher angular resolution would be required to better assess the kinematic properties in this region. The modeling of the nuclear region does not impact the analysis of the molecular wind presented in the next Section.

\section{Molecular Wind}\label{subsec:outflow}

\begin{figure}
  \resizebox{\hsize}{!}{\includegraphics{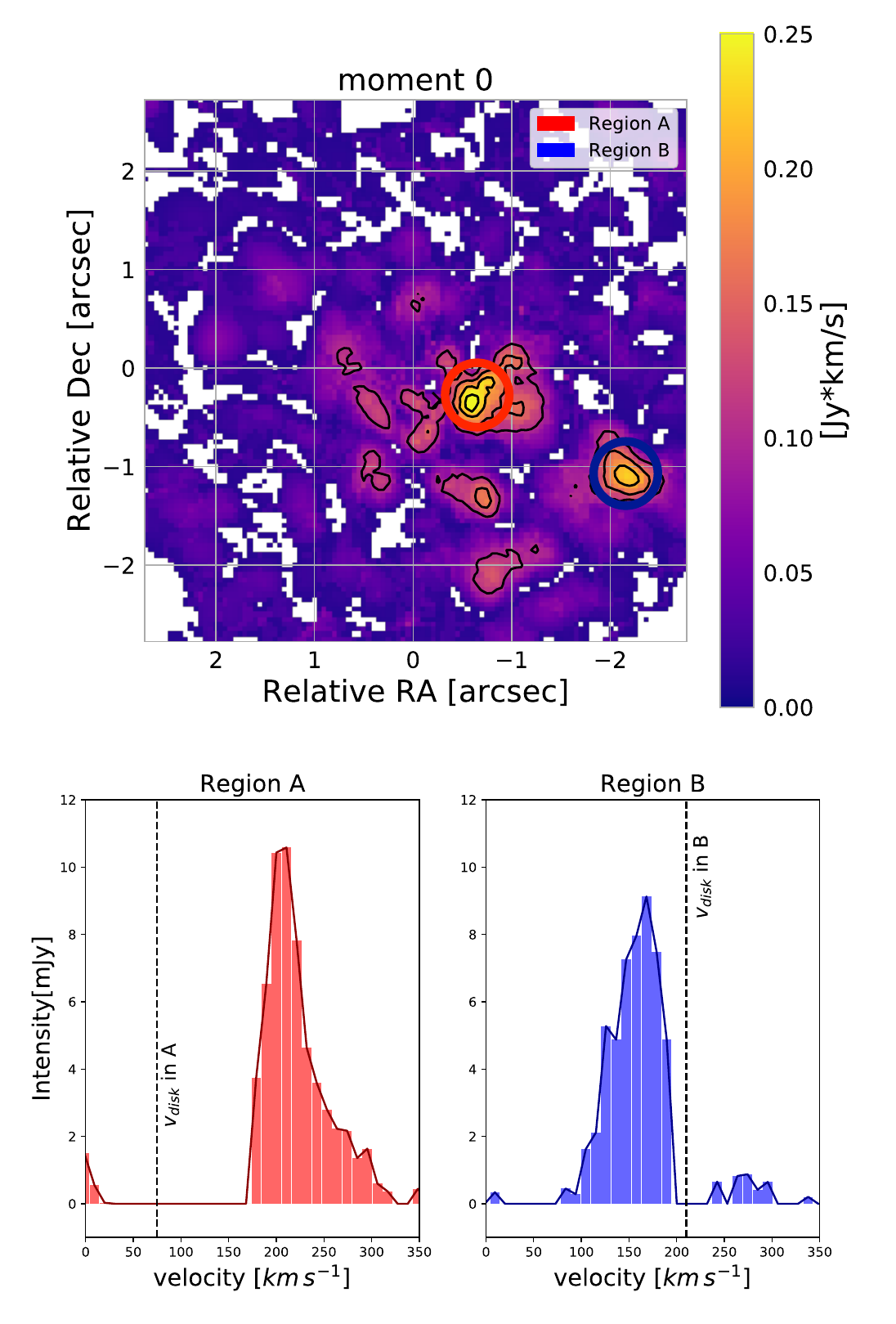}}
  \caption{Top panel: the CO(2-1) residual intensity map after the best-fit disc model has been subtracted from the data-cube. The red and blue circles indicate Regions A and B respectively. The black contours show the emission at (2, 3, 4, 5)$\times \sigma$.
  Bottom panels: spectra extracted from region A (red histogram) and B (blue histogram). The dashed black lines indicate the value of the disc rotational  velocity at A and B based on our best-fit dynamical model( $v_{disc}$ = 75 km/s around region A, and $v_{disc}$ = 210 km/s around region B). }
  \label{outflow}
\end{figure}

 Regions A and B show residuals both in the velocity and in the velocity dispersion maps.  Fig. \ref{outflow}, top panel, shows the intensity map of the residuals. The spectra extracted from A and B are shown in Fig. \ref{outflow}, bottom panels. The gas in region A has a peak velocity of 220 km/s, and a red tail reaching 300 km/s, whereas the gas in B is characterized by a peak velocity of about 170 km/s. We note that the disc rotational velocity at A and B is 70 and 210 km/s (dashed lines), respectively, thus showing that gas in these regions is not participating to the disc rotation, but it can be rather associated to a wind. The wind mass outflow rate is computed using the relation from \citet{fiore2017}, assuming a spherical sector:
\begin{equation}\label{eq1}
\dot{M}_{of} = 3 \frac{M_{of}\,v_{of}}{R_{of}}
\end{equation}
where $v_{of}=v_{98}$ is the wind velocity, the velocity enclosing 98$\%$ of the cumulative velocity distribution of the outflowing gas \citep{bischetti2019}, $M_{of}$ is the wind molecular mass, and $R_{of}$ is distance from the active nucleus reached by the wind. 
Table \ref{table:dati-outflow} gives $v_{98}$, ${M}_{of}$ and $R_{of}$ for regions A and B.  Because we do not know a-priori whether the gas in the wind is optically thin or thick, we derived the molecular mass range in A and B adopting a conversion  factor $\alpha_{CO}\,= 0.3\mathrm{\,M_{\odot} (K\,km/s\,pc^{2})^{-1}}$ for the optically thin \citep{morganti2015, dasyra2016}, and  $\alpha_{CO}\,= 0.8\mathrm{\,M_{\odot} (K\,km/s\,pc^{2})^{-1}}$ \citep{bolatto2013,morganti2015} for the optically thick case. These assumptions imply wind masses in the ranges  $M_{H_2} = (2.1-5.6) \times 10^6 \mathrm{M_{\odot}}$  and $M_{H_2} = (2.0-5.4)\times 10^6 \mathrm{M_{\odot}}$, thus a molecular outflow rate in the range $\dot{M}_{of}\,=\,5.5-14.7 \mathrm{\,M_{\odot}/yr}$ and $\dot{M}_{of}\,=\,0.86-2.30 \mathrm{\,M_{\odot}/yr}$ for region A and B respectively. The lowest boundaries are pertinent to optically thin gas and thus represent lower limits to the molecular mass and outflow rate. These estimates are derived assuming AGN-driven winds because $R_{of}$ is the distance from the AGN. Should 

We evaluated the escape velocity at A and B using the relation
\begin{equation}
v_{esc} = \sqrt{\frac{GM_{dyn}}{R_{of}}}
\end{equation}
where we used the dynamical mass derived from the best fit disc model (Fig. \ref{parameters}, right panel). We find that $v_{esc} \sim$ 140 km/s at A, that is significantly smaller than the wind velocity, suggesting that the outflowing gas can leave the innermost 300 pc region. The escape velocity at B is $\sim$ 227 km/s, similar to wind B velocity. Computing the time needed for the gas to travel from the center to the wind position and assuming a constant velocity equal to $v_{98}$, we obtain 1.16 Myr to reach A, and 6.85 Myr to reach B. Therefore these two outflowing gas components might represent two consecutive episodes of winds launched at different times, or the same event with a decelerating wind \citep[e.g.][]{fischer2013, veilleux2020}. Current data cannot discriminate between these two scenarios.

\begin{figure*}
   \centering
   \resizebox{\hsize}{!}{\includegraphics{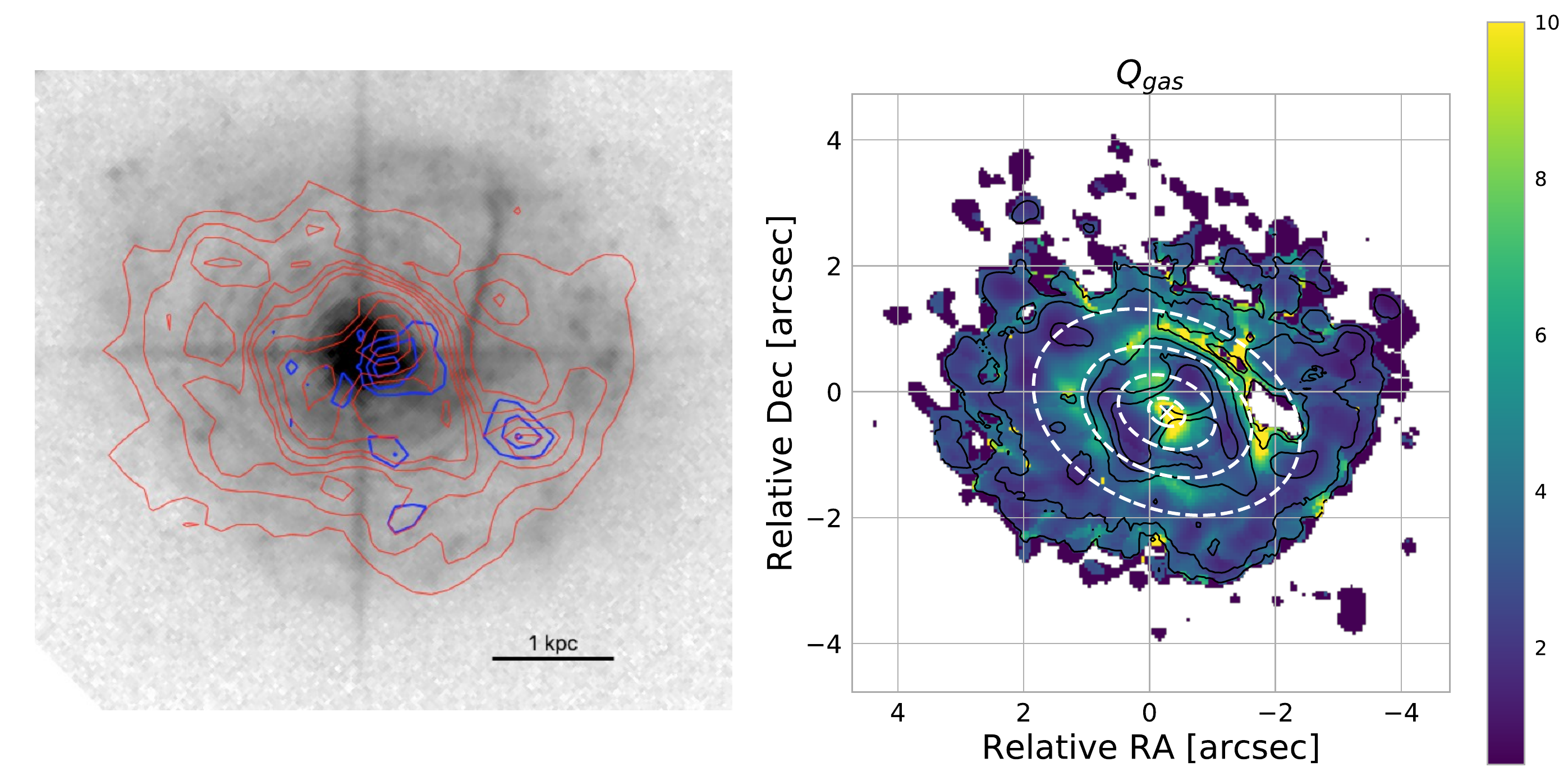}}
   \caption{Right panel: HST image in the FQ508N filter, probing rest-frame [OIII]$\lambda$5007\AA ~emission (greyscale). Red contours represent CO(2-1) emission (Figure \ref{spettro} center panel, contours are drawn at (1, 2, 3, 4, 5, 7, 10, 15, 20, 25, 30, 37)$\times \sigma$). Blue contours represent CO(2-1) emission in Region A and B (Figure \ref{outflow} top panel, contours are drawn at (3, 4, 5)$\times \sigma$). Image  from the Mikulski Archive for Space Telescopes (MAST), HST Proposal 12212. Right panel: the Toomre Q-parameter map for the cold molecular gas component. The black contours represent the moment 0 CO emission at (2, 5, 15, 30, 45)$\times \sigma$. The white cross marks the continuum peak position. The white dashed ellipses mark projected annuli with radius $R=0.3, 0.8, 1.4, 2.2$ arcsec, respectively (see Table \ref{table:toomre}).}
              \label{fig:HST-toomre}
\end{figure*}

 \begin{figure}
   \centering
   \resizebox{\hsize}{!}{\includegraphics{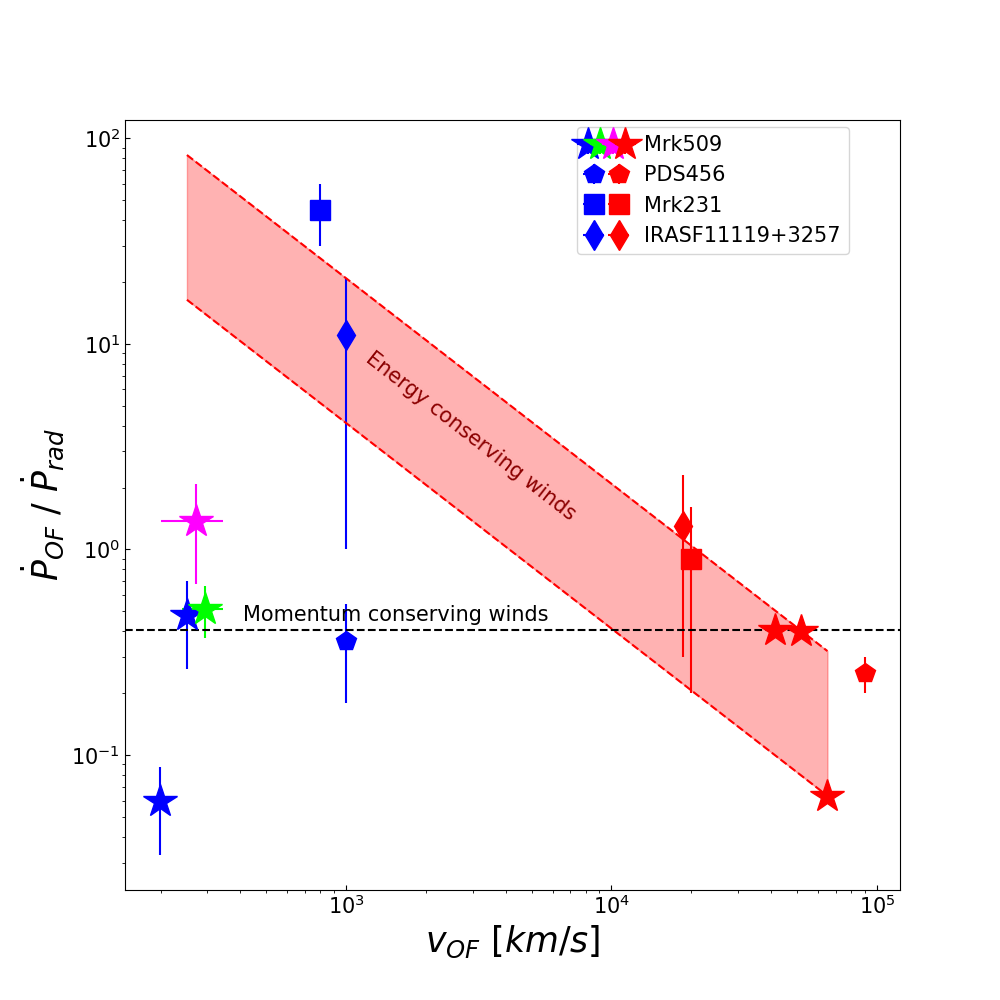}}
   \caption{The wind momentum load, $\dot{P}_{of}/\dot{P}_{rad}$, as a function of the wind velocity for Mrk 509 \citep[][and this work]{tombesi2011, tombesi2012, liu}, and a compilation of local AGN (PDS456: \citep{nardini2015, bischetti2019}, Mrk231  \citep{feruglio2015}, IRASF11119+3257  \citep{tombesi2015}. Red symbols are UFOs; the green star symbol is the ionised [OIII] wind for Mrk509; blue symbols are molecular winds, and magenta star symbol is the total wind (ionised plus molecular phases) for Mrk509. The black dashed line marks the expectation for momentum conserving wind driven by the AGN in Mrk 509. The red shaded area represents the prediction for an energy-conserving wind with $\dot{P}_{mol}/\dot{P}_{rad} = v_{UFO}/v$.} 
              \label{outfow-momentum-flux}
\end{figure}

\begin{table*}
\caption{Parameters of the molecular wind components.}    
\centering       
\label{table:dati-outflow}                        
\begin{tabular}{c c c c c c c c c}        
\hline\hline  
Region & $v_{98}$ & $\mathrm{R_{of}}$ & $S_{CO}$ & $M_{H_2}$ & $\dot{M_{of}}$ & $\dot E_{kin}$ & $\dot{P}_{of}/\dot{P}_{rad}$ & $v_{esc}$ \\
& [km/s] & [kpc] & [$\mathrm{Jy \, km/s}$] & [$10^6\mathrm{\,M_{\odot}}$] & [$\mathrm{M_{\odot}/yr}$] & [$\rm 10^{40}$ erg/s] & & [km/s]\\
& (1) & (2) & (3) & (4) & (5) & (6) & (7) & (8)\\ 
\hline
A  & 250 & 0.3 & 0.51$\pm 0.05$ & 2.1-5.6 & 5.5-14.7 & 11-30 & 0.5 $\pm$ 0.2 & 140\\
B  & 200 & 1.4 & 0.49$\pm 0.05$ & 2.0-5.4 & 0.86-2.30 & 1.1-2.9 & 0.06 $\pm$ 0.03 & 227\\
\hline                                   
\end{tabular}
 \flushleft 
\footnotesize {\bf Notes.}{(1) wind velocity $v_{98}$, (2) projected wind radius in kpc, (3) integrated CO(2-1) flux density in $\rm Jy \, km/s$ , (4) the molecular gas mass in units of $\mathrm{10^{6}}\, M_{\odot}$, (5)  mass outflow rate in $\mathrm{M_{\odot}/yr}$, (6) kinetic power of the wind in erg/s, (7) wind momentum load, and (8) escape velocity in km/s, at region A and B.}
\end{table*}

\section{Discussion}

Mrk509 is classified as a spheroid based on optical imaging \citep{Ho2014}, and it hosts a ionised gas ring and disc, where star formation is currently active \citep{kriss2011,fisher}.  
With its  SFR of $5.1 \pm 0.5 \, \rm M_{\odot}/yr$ and the stellar mass of $M_*= (1.2 \pm 0.1) \times 10^{11} \, \rm M_{\odot}$ \citep[][Appendix A]{gruppioni2016}, it sits on the Main Sequence of star-forming galaxies. Our observations show that Mrk509 hosts a molecular gas reservoir of $M_{H_2}= (1.7\pm0.1) \mathrm{\times 10^{9} \, M_{\odot}}$ within 5.2 kpc.
 The stellar mass within the same region has been derived by scaling the total stellar mass assuming an exponential profile and the size of the galaxy in K band (33 kpc, NED). It turns out $M_*(R<2.6  \mathrm{kpc})\sim 3 \times 10^{10} \rm M_\odot$. 
The molecular gas fraction $\mu = M_{H_2}/{M_*}$, in the inner $\sim 5.2$ kpc, is then $\mu \sim 5 \% $, consistent with that derived for local star-forming galaxies with the same stellar mass \citep[e.g.][]{saintonge2011, saintonge2017}.
 We find that the CO(2-1) moment 1 map shows a velocity gradient consistent with a inclined rotating disc with inclination consistent with that derived from UV imaging of the host galaxy. The derived dynamical mass is $M_{dyn}$ = (2.0 $\pm$ 1.1) $\times$ $10^{10}$ $\mathrm{M_{\odot}}$ within a $\sim$2.6 kpc radius. The uncertainty is dominated by the uncertainty on the inclination. The dynamical mass is thus consistent with the $M_*(R<2.6\mathrm{kpc})$. 
 Our dynamical modeling also suggests a possible warp in the disc towards the center, as reported by  \citet{combes2020} in other nearby Seyfert galaxies. This nuclear region, however, remains marginally resolved in current observations, and we cannot exclude more complex structures, such as bars or rings.\\
 The molecular disc kinematics and inclination are consistent with those reported for the ionised gas phase by \citet{fisher} and \citet{liu}, who mapped the ionised disc on scales of up to 4 kpc, see Fig. 6 in \citet{liu}.
We compare our CO(2-1) moment 0 map with the HST image in FQ508N filter, which traces ionised gas emitting in the [OIII]$\lambda$5007\AA, to investigate the relation between cold molecular and ionised gas distributions. Figure \ref{fig:HST-toomre} left panel shows that the molecular gas (red contours) is detected  at the position of the starburst ring. 
The [OIII] image shows a linear tail extending out from the starburst ring toward north-west. The "jut" of tail that seems to bend towards the nucleus is also visible. These two structures have been interpreted as due to a minor merger with a gas-rich dwarf galaxy \citep{fisher}. In our ALMA data we do not detect CO-emitting gas at the position and velocity of the linear tail. 
The 3$\sigma$ upper limit on the $H_2$ mass in the tail is $<10^7$ $\rm M_{\odot}$, based on the r.m.s. of our observation, adopting a Milky Way conversion factor and FWHM = 50 km/s for the CO line \citep[see][]{aalto2001}. The limit appears consistent with a minor merger scenario \citep{knierman2013, vandevoort2018}.
Both the tidal feature and the presence of a molecular disc with an ongoing starburst are in agreement with the scenario where galaxy interactions and mergers produce gas destabilization, and feed both star-formation and AGN  activity \citep[e.g.][and refs therein]{ menci2014} . 
Assuming that most SFR occurs within the starburst ring, we derive a star-formation efficiency $SFE = SFR/M_{H_2} =  6.2 \times 10^{-9} \rm yr^{-1}$, consistent with that found for local early-type galaxies with recent minor mergers \citep{saintonge2012, davis2015}. 
\\We quantify the disc stability against gravitational fragmentation, by deriving a map of the Toomre Q-parameter \citep{toomre1964} for the cold molecular gas component, $Q_{gas}$. 
 The Toomre parameter is related to the local gas velocity dispersion $\sigma_{disp}$, the circular velocity $v_{rot}$, and the gas surface density $\Sigma_{gas}$ at any given radius R, through the relation
\begin{equation}
Q_{gas} = \frac{a v_{rot} \sigma_{disp}}{\pi R G \Sigma_{gas}},
\end{equation}
where $a$ is a constant that varies from 1 (for a Keplerian rotation curve), 1.4 (for a flat rotation curve), to 2 (for a solid-body rotation curve), see \citet[][]{genzel2011,genzel2014} and refs therein. 
To obtain a map of $Q_{gas}$ we use $v_{rot}$, derived through the $^{3D}BAROLO$ model of the whole disc, using the assumed inclination and best fit PA. We adopt $a$=1 up to $R$=1.1 arcsec, where the rotation curve is steep, and $a$=1.4 at larger radii. The velocity dispersion and $\Sigma_{gas}$ are calculated pixel by pixel from the moment 2 and 0 maps, respectively; the radius $R$ is deprojected.
The map of $Q_{gas}$ is shown in Fig. \ref{fig:HST-toomre} right panel, while in Table \ref{table:toomre} we report $\sigma_{disp}$, $\Sigma_{gas}$, and $Q_{gas}$ in 5 annulii between $R$=0.3 and 0.8 arcsec, 0.8 and 1.4 arcsec, 1.4 and 2.2 arcsec (that includes a ring-like feature of enhanced $\sigma_{disp}$ visible in Fig. 2), $R>$ 2.2 arcsec, and in the whole disc excluding the central region of 0.3 arcsec radius, to avoid the beam-smearing effect.
We find values that range from $Q_{gas}\approx$ 0.5 up to 10 and we note that $Q_{gas}$ exceeds 1 across most of the molecular disc.  
A similar $Q_{gas}$ range  was reported by e.g. \citet[][]{garciaburillo2003, sani2012} in nearby Seyfert galaxies, and \citet{hitschfeld2009} in M51.  
Where the value of the Q-parameter is higher than the critical value $Q_{crit}$ the disc is unstable against gravitational collapse. The commonly adopted critical value of $Q_{gas}$ is in the range 1-3 \citep{genzel2014, leroy2008}, whereas if one considers the total $Q_{tot}^{-1} = Q_{gas}^{-1} + Q_{star}^{-1}$, including both the gas and stellar contribution, the critical value is $\sim1$ \citep{aumer2010}. The $Q_{gas}$ we find in Mrk 509 implies that the disc is unstable in its outer regions, across the starburst ring, whereas is stable against fragmentation at nucleus and in a lopsided ring-like structure located at $R\sim2$ arcsec from the AGN.
\begin{table}
\caption{Average $Q_{gas}$ and velocity dispersion in 4 annulii.}    
\centering       
\label{table:toomre}                        
\begin{tabular}{c c c c}        
\hline\hline  
Annulus & $\sigma_{disp}$ & $\Sigma_{gas}$ & $Q_{gas}$\\
 & [km/s] & [$\rm M_{\odot}/pc^2$] & \\
(1) & (2) & (3) & (4)\\ 
\hline
0.3 $ < R < $ 0.8  & 46.1 & 475 & 5.0 \\
0.8 $ < R < $ 1.4  & 28.4 & 318 & 3.5 \\
1.4 $ < R < $ 2.2  & 18.3 & 118 & 4.8 \\
$ R > $ 2.2  & 7.6 & 42 & 2.4 \\
$ R > $ 0.3  & 12.6 & 94 & 3.0 \\
\hline                                   
\end{tabular}
 \flushleft 
\footnotesize {\bf Notes.}{(1) Annulus region with deprojected radius $R$ in arcsec, (2) average velocity dispersion, (3) average gas mass surface density in $\rm M_{\odot}/pc^2$ (4) average Toomre parameter.}
\end{table}

\begin{figure}
   \centering
   \resizebox{\hsize}{!}{\includegraphics{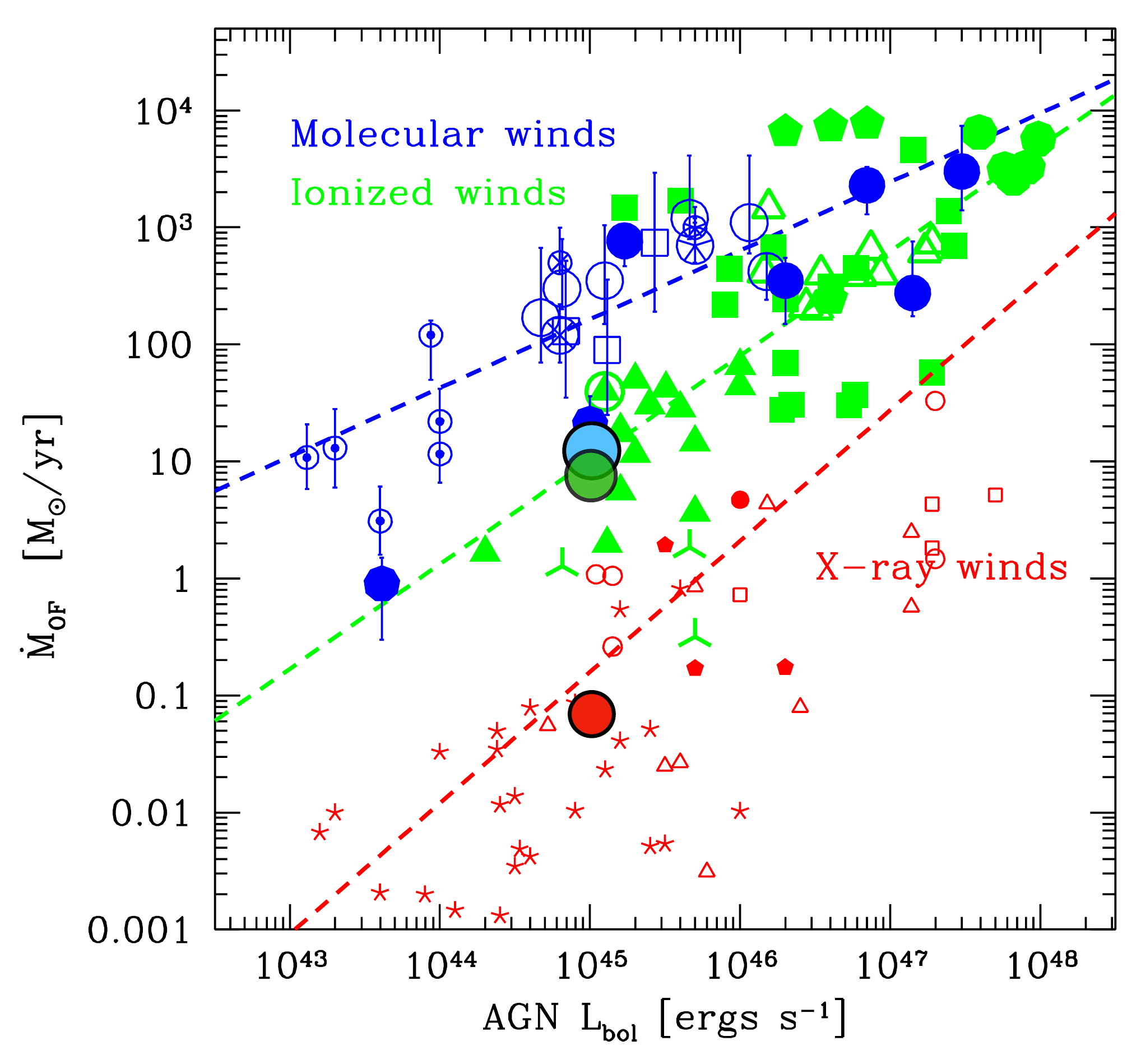}}
   \caption{The wind mass outflow rate versus AGN bolometric luminosity for Mrk509 and a compilation of AGN from \citep{fiore2017}. For Mrk 509 we plot the $\dot M_{of}$ of the total molecular wind (this work, light blue circle), the ionised wind \citep[green circle]{liu}, and the UFO \citep[red circle]{tombesi2010,tombesi2011,tombesi2012}. The size of the circles represents the uncertainty in $\dot M_{of}$. Small blue symbols: molecular gas winds. Small green symbols: ionised gas winds. Small red symbols: highly ionised gas winds detected in X-rays. The dashed blue, green and red lines are the best fit correlations of the molecular, ionised and X-ray absorbers samples, respectively. Adapted from \citet{fiore2017}.}
              \label{mdotlbol}
\end{figure}

\noindent We find significant kinematic perturbations at different locations across the disc, where the molecular gas shows deviations from the disc rotation pattern. The first (Region A) is located at a distance of $\sim$300 pc from the nucleus, the second (Region B) further out at a distance of about 1.4 kpc. 
 In the proximity of Region B we detect a CO-depleted region (Fig. \ref{outflow}). This is close (in projection) to the location where the linear tail related to the merger event observed with HST crosses the disc. The CO-depleted region may be related to the merger event locally decreasing the CO in the disc, however the optical image does not show any lack of ionised gas there.
In Section \ref{subsec:outflow}, we showed that both A and B regions show non-circular gas motions, that are likely due to outflowing gas.  
Regarding Region A, integral field spectroscopy (IFS) spectroscopy shows a [OIII] wind with velocity similar \citep{liu} to the molecular one, supporting the notion that this region hosts a multiphase wind. The ionised and molecular winds are co-spatial, within the accuracy allowed by the angular resolution of the data, suggesting a cooling sequence \citep{richings2017,menci2019}.
Region B has a small projected distance from the tidal tail seen in [OIII]. The ionised gas velocity in the tidal tail is $-100$ km/s, according to \citet{liu}, while it is 200 km/s for the molecular gas in B, and 250 km/s in A. Therefore it is unlikely that the molecular gas in A and B is related to the tidal tail. \\
It is likely that molecular gas in A and B is involved in an AGN-driven wind, whose total molecular outflow rate is in the range 6.4 - 17.0 $\mathrm{M_\odot/yr}$, for the optically thin and thick cases respectively. 
The contribution of Region B to the global outflow budget is a factor of $\sim$10 smaller than Region A, therefore including or excluding region B, does not change the main result of the work.
We computed the wind momentum load, that is the ratio between the outflow momentum flux, $\dot{P}_{of} = \dot{M}_{of} \cdot v_{98}$, and the radiative momentum flux, $\dot{P}_{rad} = L_{bol}/c$. We adopt a bolometric luminosity $L_{Bol}=  10^{44.99}\mathrm{\,erg\,s^{-1}}$ derived by \citet{duras2020} from SED fitting and consistent with that derived by \citet{gruppioni2016}. The wind momentum load values obtained are reported in Fig. \ref{outfow-momentum-flux} for the different wind components: the three Ultra-fast Ouflow (UFO) components from \citet{tombesi2010,tombesi2011,tombesi2012}, the ionised gas winds from \citet{liu}, and the molecular winds from this work. The molecular wind A and the ionised gas wind show similar velocity and arise from approximately the same region, and also show similar mass outflow rates, suggesting that they might be part of a multiphase wind. The molecular wind A and the ionised wind are consistent with a momentum conserving driving (black dashed line), and are well below the expectation for energy conserving winds. Wind B has a much lower mass outflow rate. From current data we cannot discriminate if A and B components of the molecular wind are due to consecutive episodes or to the same event. Assuming that both the molecular components and the ionised wind detected by \citet{liu} are part of the same multiphase wind, we derive a total mass outflow rate, molecular plus ionised, of $\sim 22 \,\mathrm{M_\odot/yr}$. For comparison, the accretion rate onto the SMBH is $\dot{M}_{acc} = L_{Bol}/\epsilon c^2 \sim 0.176 \, \mathrm{M_{\odot}/yr}$, assuming an efficiency $\epsilon = 10 \%$.
The total wind momentum load, $\dot{P}_{OF}/\dot{P}_{rad} \sim$ 1.4 $\pm$ 0.6 (magenta point) is below the energy conserving wind expectation, and compatible with a momentum conserving wind powered by the nuclear semi-relativistic wind probed by the UFOs. These momentum boost can also be attained by radiation pressure on dust \citep{ishibashi2021}.
The wind mass-loading factor is $\eta = \dot{M}_{of}/SFR=4.4$, in agreement with what found for molecular winds in other Seyfert galaxies \citep[e.g.][] {fiore2017}. 
\\ In the following we discuss the possibility that the outflows are driven by energy injected by supernova (SN) explosion occurring in the starburst ring. To derive the mass outflow rate at B in this case, we adopt a radius $R= 0.4$ kpc, that is half the projected size of the starburst ring. This implies $\dot M=3-8 \, \rm M_{\odot}/yr$ by adopting equation \ref{eq1}, and $\dot P= 0.4-1.0 \times 10^{41} \rm erg/s$.
Assuming that the total SFR (5 $\rm M_{\odot}/yr$) is homogeneously distributed across the ring, we consider a portion of the starburst ring equal to 1 tenth of the ring in area, and
we estimate a kinetic power of a SN-driven wind $\dot E_{kin,SN}= 3.5 \times 10^{41}$ erg/s, assuming a SN rate of 0.02($\rm SFR/M_{\odot}/yr$) \citep[e.g.][]{veilleux2005}. 
Outflow B is then energetically consistent with a SN-driven wind originating from the starburst ring \citep{venturi2018, leaman2019}.
Regarding wind A, the SFR at its location cannot be resolved in current data, but it is most likely a small fraction of the total SFR which occurs mainly in the starburst ring. Therefore outflow A is most likely driven by the AGN.
\\ Figure \ref{mdotlbol} shows the outflow rate, $\dot M_{of}$, versus the AGN bolometric luminosity for Mrk 509 and a compilation of AGN adapted from \citet{fiore2017}. Mrk 509 shows a ionised wind and X-ray winds that are broadly consistent with the correlations found by \citet{fiore2017}, whereas its molecular outflow rate is significantly below the best-fit correlation of molecular winds. Indeed, the 
ionised outflow rate is similar to the molecular one, unlike the average AGN with similar $L_{bol}$. We note, however, that the correlation is derived from a biased sample which collect mainly high luminosity objects and known massive winds, and therefore is probably biased towards positive detection, boosting the outflow rates. This highlights the importance of performing statistical studies on unbiased AGN samples to derive meaningful scaling relations. This study will be carried out by using the whole IBISCO sample in a separate publication. 

%

  
%

%
   

\section{Conclusions}

We presented an analysis of the CO(2-1) line and 1.2 mm continuum of Mrk509, a Seyfert 1.5 galaxy drawn from the IBISCO sample of hard-X-ray selected local AGN. Mrk509 is optically classified as a spheroid and hosts ionised winds in different gas phases, from warm ionised gas wind traced by [OIII], to highly ionised UFOs seen in X-rays. We report the following findings:
\begin{enumerate}
    \item The galaxy hosts a molecular gas reservoir of $M_{H_2} = (1.7\pm0.1), \times 10^9\, \mathrm{M_{\odot}}$ within the inner 5.2 kpc, and partially overlapping with the starburst ring seen in HST optical imaging. In this region we estimate a molecular gas fraction $\mu \sim 5 \%$, consistent with that of local star-forming galaxies with the same stellar mass \citep[e.g.][]{saintonge2011, saintonge2017}.  
    Both the signatures of a minor merger and the presence of a molecular disc with an ongoing starburst are in agreement with the scenario where galaxy interactions and mergers produce gas destabilization, and feed both star-formation and AGN activity. 
    The star formation efficiency across the molecular disc and starburst ring, $SFE  = 6.2 \times 10^{-9} \rm yr^{-1}$, is consistent with that found for local early-type galaxies with recent minor mergers.
    \item We quantify the disc stability by estimating the spatially-resolved Toomre Q-parameter for the cold molecular gas component, $Q_{gas}$. We find that $Q_{gas}$ varies across the molecular disc in the range $\sim0.5-10$. The disk is unstable across the starburst ring, and stable against fragmentation at nucleus, in a lopsided ring-like structure located at $R\sim2$ arcsec from the AGN, and at the location of the molecular winds.
    \item The main velocity gradient detected in CO(2-1) is well modeled by a disc with $M_{dyn}$ = (2.0$\pm$1.1) $\times$ $10^{10}$ $\mathrm{M_{\odot}}$ within $\sim$ 5.2 kpc and inclination $44\pm10$ deg with respect to the line of sight. The CO kinematics in the nuclear region, r$\sim700$ pc, is barely resolved at the current angular resolution and may be  affected by a warped nuclear disc. Higher angular resolution observations may help to better constrain the gas distribution and kinematics in the nuclear region. 
    \item   We find significant perturbations of the molecular gas kinematics at two different locations in the disc, where the molecular gas shows deviations from the disc rotation, that we interpret as molecular winds. The wind A has a velocity of $v_{98}=250$ km/s and is located at a distance of $\sim$300 pc from the AGN, in the same region where a ionised gas wind is detected. 
    The wind B is found at a distance of about 1.4 kpc, overlapping on the line of sight with the starburst ring, and at a small projected distance from the tidal tail. Its velocity $v_{98}=200$ km/s suggests that B is not related to the tidal tail. 
    The total molecular outflow rate for A and B is in the range $6.4 - 17.0$ $\mathrm{M_\odot/yr}$, for the optically thin and thick cases respectively, with outflow B constituting only a small fraction of the total outflow rate (about 10 $\%$). 
    The wind kinetic energy is consistent with a momentum-conserving wind driven by the AGN. 
    The spatial overlap of the ionised and molecular winds in A, and their similar velocity, suggest a cooling sequence within a multi-phase AGN-driven wind.
    Whereas outflow B is consistent with a SN-driven wind originating in the starburst ring.
\end{enumerate}

\begin{acknowledgements}
We thank the referee for his/her careful reading and insightful suggestions that helped improving the paper.
This paper makes use of the ALMA data from project: ADS/JAO.ALMA$\#$2017.1.01439.S. ALMA is a partnership of ESO (representing its member states), NSF (USA) and NINS (Japan), together with NRC (Canada), MOST and ASIAA (Taiwan), and KASI (Republic of Korea), in cooperation with the Republic of Chile. The Joint ALMA Observatory is operated by ESO, AUI/NRAO and NAOJ. The National Radio Astronomy Observatory is a facility of the National Science Foundation operated under cooperative agreement by Associated Universities, Inc. We acknowledge financial support from PRIN MIUR contract 2017PH3WAT, and PRIN MAIN STREAM INAF "Black hole winds and the baryon cycle".
\end{acknowledgements}

%
%
\bibliographystyle{aa} 
\bibliography{bibliografia} 
\begin{appendix}
\section{Stellar Mass}\label{appendixa}
\begin{figure}
\resizebox{\hsize}{!}{\includegraphics{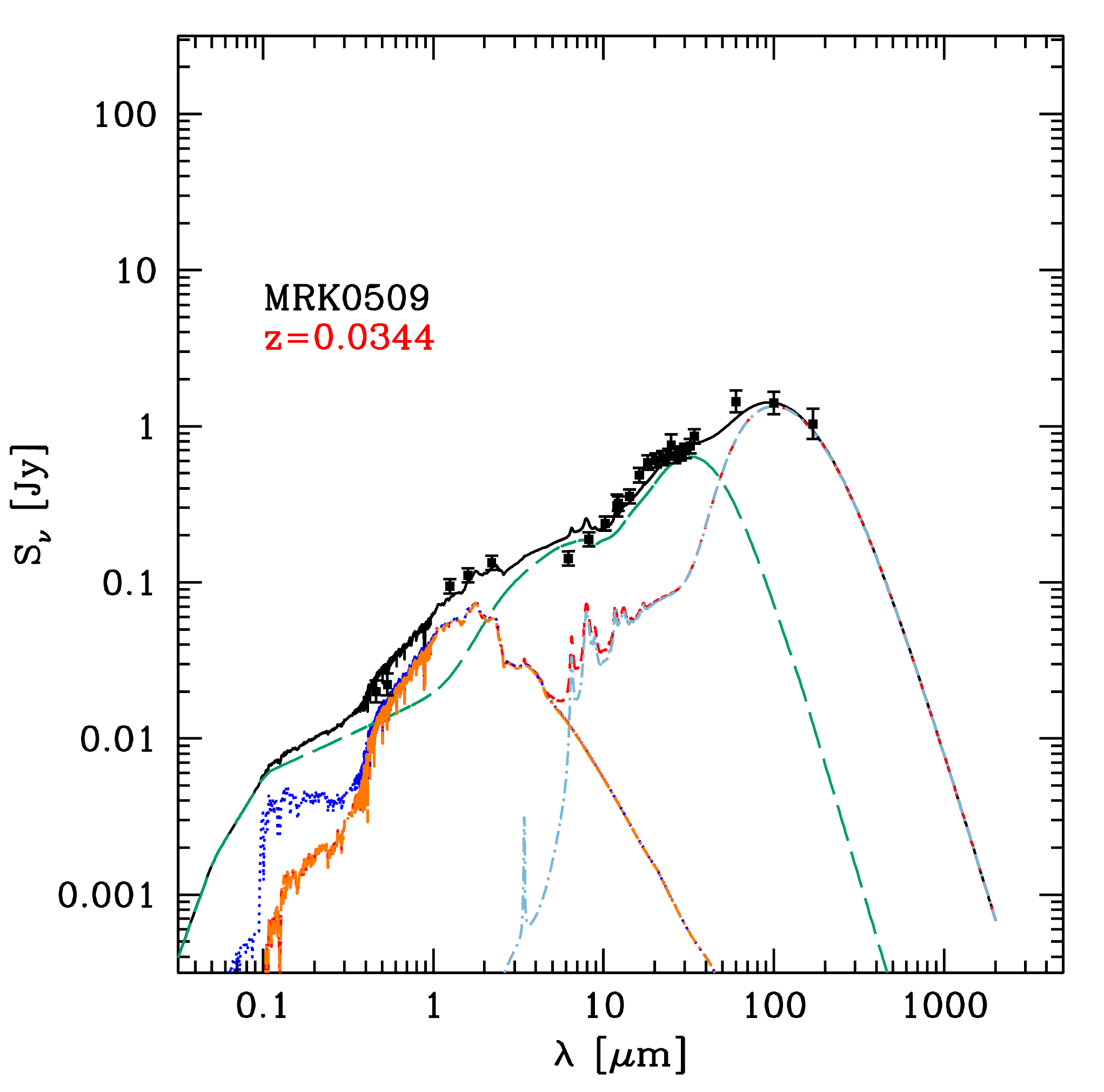}}
\caption{Mrk509 observed SED decomposed into stellar, AGN and SF components using the technique described by \citet{gruppioni2016}. The black filled circles with error bars are data, blue dotted lines show the unabsorbed stellar component, red dashed lines show the combination of extinguished stars and dust IR emission, while long-dashed green lines show the dusty torus emission. Pale-blue dot–dashed lines show the dust re-emission, while black solid lines are the sum of all components (total emission).}
\label{fig:appenddixa}
\end{figure}

The rest-frame SED is fitted with a three components model using the approach described in \citet{gruppioni2016}. The adopted code combines three components simultaneously, i.e. the stellar component, the dusty star-formation part and the AGN/torus emission. 
The adopted libraries are the \citet{bruzual} stellar library, the \citet{dacunha2008} IR dust-emission library and the library of AGN tori by \citet{fritz2006}, updated by \citet{feltre2012}. The latter includes both the emission of the dusty torus, heated by the central AGN engine, and the emission of the accretion disc. The considered AGN models are based on a continuous distribution of dust across the torus. 
The torus emission is shown as green dashed line. The decomposition code provides a probability distribution function (PDF) in each photometric band and for each fit component, allowing an estimate of the uncertainty related to each decomposed contribution. In Fig. \ref{fig:appenddixa}, we show the observed SED of Mrk509 with the galaxy and AGN components. We defer to \citet{gruppioni2016} for further details about the SED decomposition method.  This SED decomposition delivers the star-formation rate $SFR= 5.1 \pm 0.5 \, \rm M_{\odot}/yr$  \citep{gruppioni2016}, and the stellar mass $M_*= (1.2 \pm 0.1) \times 10^{11} \, \rm M_{\odot}$ (this work). 

\section{Inner disc model}\label{appendixb}

\begin{figure*}
\resizebox{\hsize}{!}{\includegraphics{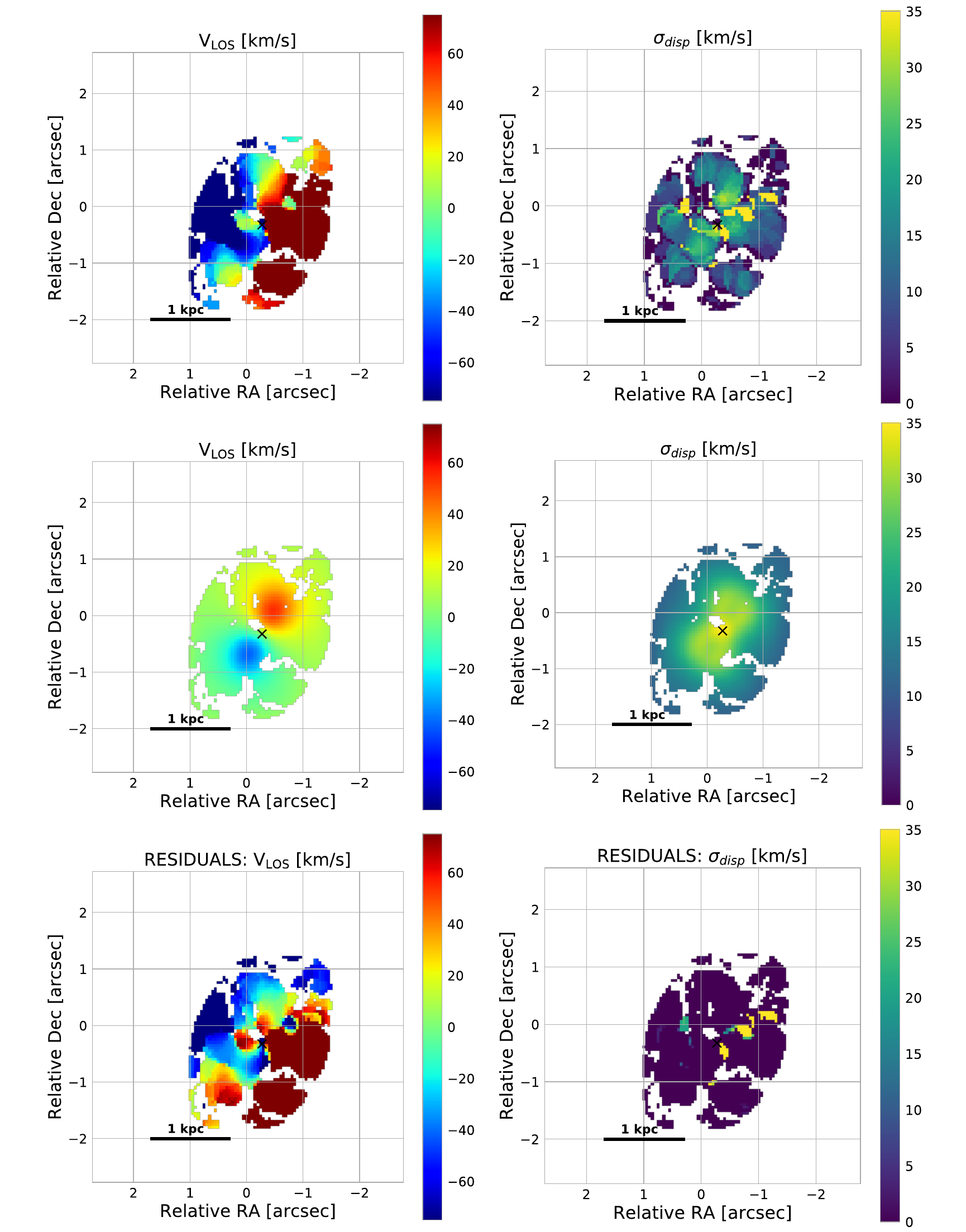}}
\caption{Top panels: the CO(2-1) moment 1 (top-left) and moment 2 maps (top-right) computed from the data-cube, after subtraction of the disc model (Section \ref{sec:results}). A threshold of $3\sigma $ has been applied. The black cross marks the AGN position. Central panels: the moment 1 (central-left) and moment 2 (central-right) maps of the disc model of the inner r=700 pc region. Bottom panels: the moment 1 (bottom-left) and moment 2 (bottom-right) maps of the residuals obtained by subtracting the model velocity and velocity dispersion maps (central panels) from the observed ones (top panels).}
\label{fig:appendixb}
\end{figure*}

 Based on the residual maps of the disc model (Section \ref{sec:results}), we perform a dynamical modeling of the inner $\sim$ 1.5 kpc region, where the model of the large scale disc produces significant blue- and red-shifted residuals (Fig. \ref{data-model-res}). This nuclear region is barely resolved with the current angular resolution but we attempt to model it with a nuclear disc using the same technique as explained in Section \ref{sec:results}. Given the limited signal-to-noise ratio and angular resolution we fix the inclination to 66 deg, derived from the minor/major axis ratio of the inner emitting region measured from the zeroth moment of the emission line data-cube. We fix the PA value to  330 deg, i.e. across the two brightest emission regions in the residual intensity map, Fig. \ref{residuo0}. 
 We allow two parameters to vary: the rotational velocity $v_{rot}$ and the velocity dispersion $\sigma_{gas}$ with first guess values set to 80 and 50 km/s, respectively (Fig. \ref{residuo0} right panel). The moment 1 and 2 maps of this model are reported in Fig. \ref{fig:appendixb} central panels. The right panel of Fig.\ref{residuo0} shows the position velocity diagrams cut along the kinematic major axis (330 deg); where the grey-scale and blue contours refer to the data and the red contours to the model.

\end{appendix}

\end{document}